\documentclass[aps,prl,preprint]{revtex4-1}
\usepackage{graphicx}
\usepackage{amsmath}
\usepackage{amssymb}
\usepackage{color}

\begin{document}
\title{Simulations of Coulomb systems with slab geometry using an efficient 3d Ewald summation method}

\author{Alexandre P. dos Santos}
\email{alexandre.pereira@ufrgs.br}
\affiliation{Instituto de F\'isica, Universidade Federal do Rio Grande do Sul, Caixa Postal 15051, CEP 91501-970, Porto Alegre, RS, Brazil}

\author{Matheus Girotto}
\email{matheus.girotto@ufrgs.br}

\affiliation{Instituto de F\'isica, Universidade Federal do Rio Grande do Sul, Caixa Postal 15051, CEP 91501-970, Porto Alegre, RS, Brazil}

\author{Yan Levin}
\email{levin@if.ufrgs.br}
\affiliation{Instituto de F\'isica, Universidade Federal do Rio Grande do Sul, Caixa Postal 15051, CEP 91501-970, Porto Alegre, RS, Brazil}

\begin{abstract}
We present a new approach to efficiently simulate electrolytes confined between infinite charged walls using a 3d Ewald summation method. The optimal performance is
achieved by separating the electrostatic potential produced by the  charged walls from the electrostatic potential of electrolyte. 
The electric field produced by 3d periodic images of the plates is constant, with the field produced by the transverse images of the charged plates canceling out. We show that under suitable renormalization, the non-neutral electrolyte confined between charged plates can be simulated using 3d Ewald summation with a correction that accounts for the conditional convergence of the resulting lattice sum. The new algorithm is at least an order of
magnitude more rapid than the usual simulation methods for the slab geometry and can be further sped up by adopting Particle--Particle Particle--Mesh ($P^3M$) approach.
\end{abstract}

\maketitle

\section{Introduction}

Study of electrolyte solutions is of paramount importance in physics, chemistry, and biology. Electrolytes are fundamental to human physiology~\cite{FiSe61}, but also  play an important role in systems as distinct as water soluble paints~\cite{KoAk07}, cement~\cite{UcHa97}, supercapacitors~\cite{LeGo99,WiBr04}, etc. The long range nature of the Coulomb force  makes it very difficult to obtain quantitative understanding of these systems. The well known Poisson-Boltzmann~(PB) equation can provide valuable insights for weakly interacting Coulomb systems for which electrostatic correlations are negligible~\cite{Le02}. However, many interesting phenomena, such as like-charge attraction~\cite{LiLo99,DiTa99,HaLu10,SaTr11,MaIb11} and charge reversal~\cite{Pa80,GuJo84,LeHo08,DoDi10}, appear when PB equation looses its validity. To study such systems a number of theoretical approaches have been introduced. These fall into three main categories: integral equations~\cite{Pa80,KjMa86,CoDo12}, field theory~\cite{Ne01,MoNe02}, and density functional theory~\cite{DiTa99,HeLa11}. All of these methods, however, rely on approximations which must be tested ``experimentally".  The only ``exact" quantitative approach for studying 3d Coulomb systems relies on Molecular Dynamics (MD) or Monte Carlo~(MC) simulations~\cite{AlTi87}. Unfortunately, because of the long range interaction, simulations of Coulomb systems are notoriously challenging. The difficulty arises because unlike for systems with short range forces, one can not use periodic boundary conditions for the simulation box.  Instead, an infinity of periodic replicas of the simulation cell must be constructed. Each ion in the principal simulation cell interacts with an infinite number of images of all the other ions. In order to efficiently sum over the replicas, Ewald summation methods have been  developed~\cite{Ew21,DaYo93,EsPe95}. These methods rely on splitting the interaction potential into short and long range contributions, so that the short range part can be rapidly calculated in the real space, while the long range part can be efficiently summed in the reciprocal, Fourier, space. Ewald summation methods are particularly useful for 3d isotropic systems.  However, when a system has a reduced 2d symmetry, application of Ewald summation techniques becomes more challenging. The difficulty in these cases is the appearance of Bessel function in 2d Fourier transforms, contrary to a simple exponential present in 3d, leading to a very slow convergence~\cite{Le91,WiAd97}. This problem not withstanding, there is a great practical importance to understand 
systems with reduced symmetry. These relate to the class of problems with characteristic slab geometry -- water and ionic liquids confined in thin films~\cite{RaLa01,Ma05nature,JiJa15}, charged nanopores~\cite{WoMu07,CaHa14,Bu15}, self-assembled monolayers~\cite{HuBa97}, polymer layers~\cite{HaMa00}, heterogeneous charged surfaces~\cite{NaPo05,SiBe12,BaDo15}, just to cite a few examples. 

The efficiency of Ewald-like 2d and 1d methods is not nearly as high as for isotropic 3d systems.  The slow convergence rate  was the subject of extensive studies~\cite{Ma05,Sp94}. A number of different approaches have been tried to overcome this difficulty~\cite{HaKl92,Sp97,YeBe99,KaMi01,ArDe02,ArHo02}. 
In the present paper we will introduce a new method to simulate electrolyte solutions confined by the charged walls. To avoid the slow convergence of 2d Ewald approach, we will use 3d Ewald summation.  This means that the system will be replicated in all three dimensions.  In reality, however, we are only interested in  2d ($x,y$) part of the replication, with the transverse $z$-replicas being an artifact of the 3d Ewald summation.  To diminish the effect of $z$-replicas, we will include a vacuum region
on both sides of the slab within the simulation cell.  
This, however, is not sufficient to adopt  3d Ewald
summation to 2d geometry.  The conditional convergence of the lattice sum 
still results in a surface contribution to the total electrostatic energy
which depends on the aspect ratio of the macroscopic system (sum of all the replicas).  Since we are interested in an infinite 
slab, the aspect ratio should be such that the $x$ and $y$ sides of the 
slab are infinitely bigger than the slab width ($z$-direction).  For a 
conditionally convergent lattice sum 
this means that the summation has to be first done over 
$x$ and $y$ directions, and then over $z$-direction.  This important point was discussed by Smith~\cite{Sm81} and implemented in simulations by Yeh and Berkowitz~\cite{YeBe99}~(YB). The approach of YB is quite simple. If the system
consists of electrolyte and charged plates, one can discretize the surface charge
and apply 3d Ewald summation method, with an additional surface correction, to the whole system, i.e. electrolyte and the wall charges.  
Clearly this is not very efficient since it requires to include in the lattice sum the surface charges which are fixed throughout the simulation.
Since the electric field produced by the plates is constant, it should be possible
to separate it from the rest of the system, allowing the ions to move in
a fixed external potential produced by the plates, which has a simple linear form. 
The difficulty with this approach
is that a system of only ions, without the wall charges,  is not charge neutral, so that the lattice sum will diverge.  In this paper
we will show, however, 
that this divergence can be renormalized away, allowing us to construct 
a very fast and efficient algorithm for simulating ionic systems in a slab geometry.

\section{Method}

\begin{figure}[h]
\begin{center}
\includegraphics[scale=0.25]{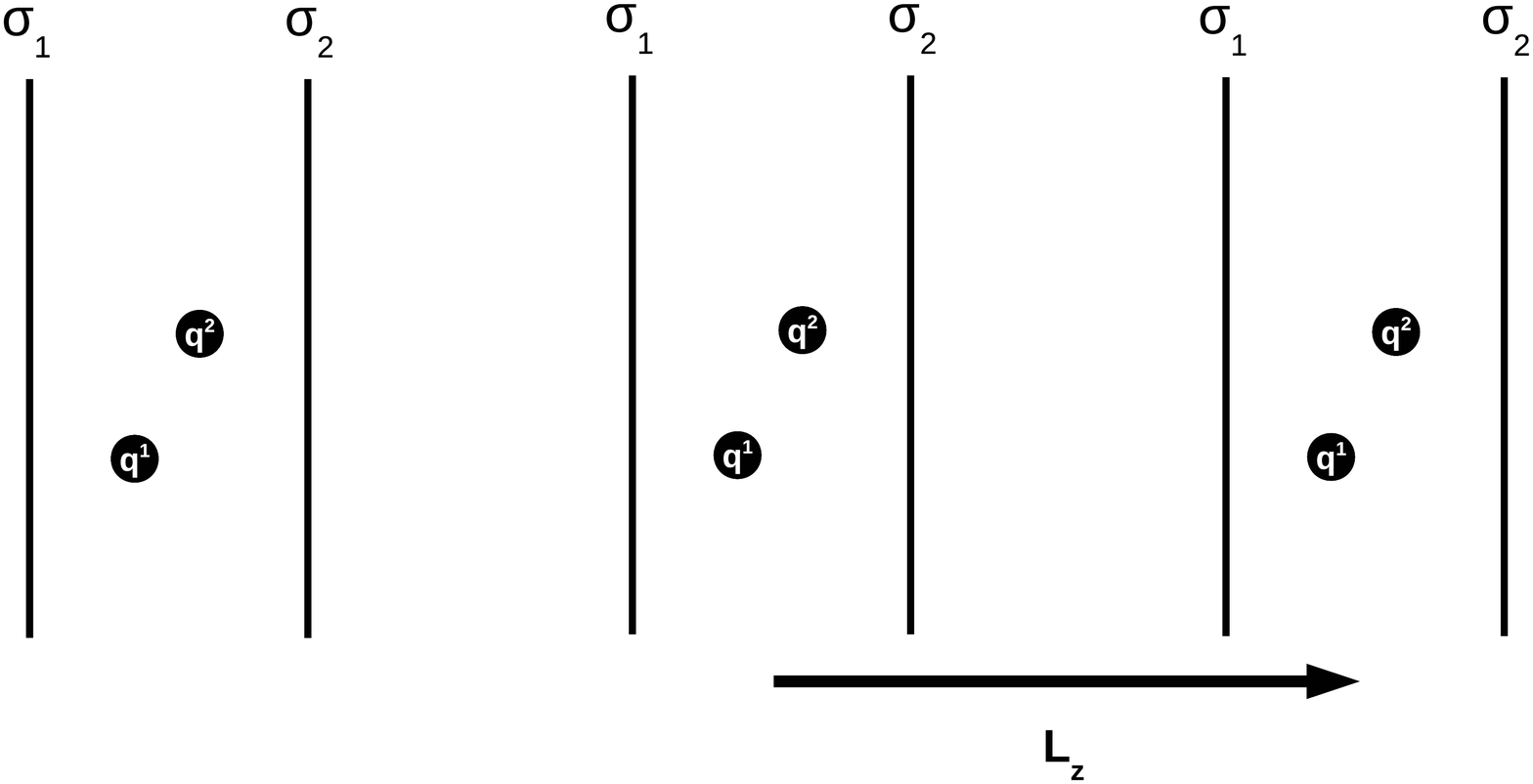}
\end{center}
\caption{3d replicated system. Note that inside the central simulation cell
the electric field produced by the $z$-replication of charged walls cancels out. }
\label{fig1}
\end{figure}

The idea of the present method is to consider the electrostatic potential produced by the plates as an external scalar field acting on all the ions inside the simulation cell. As we intend to use the 3d Ewald  summation to accelerate the simulations, we must consider the replicas of the plates in $z$-direction in addition to the replicas in $x$ and $y$-directions. The electric fields of two infinite uniformly charged plates are $2\pi\sigma_1/\epsilon_w$ and $2\pi\sigma_2/\epsilon_w$, where $\sigma_1$ is the charge density of the left plate and $\sigma_2$ of the right plate, and $\epsilon_w$ is the dielectric constant of the medium, normally water. Both fields are orthogonal to the  plates. The replication of the simulation cell in the $x$ and $y$ directions
will naturally result in 2 infinite plates. However, the replication of the simulation cell in the $z$ direction will produce an infinite array of such infinite surfaces, see Fig.~\ref{fig1}.
We note, however, that the electric fields that these $z$-images of the plates  produce on the ions
inside the simulation box cancels out, so that the ions in the cell feel only
the electric fields of the bounding walls and of their $x$ and $y$ replicas.  These
are precisely the electric fields of the infinite charged plates: $2\pi\sigma_1/\epsilon_w$ and $2\pi\sigma_2/\epsilon_w$. We can, therefore,
separate the electric field (or equivalently the electrostatic potential) 
produced by the charged plates and their images from the field produced by 
the ions and their images.
For different macro-charged bodies such as a nanopore, we cannot proceed in this way, there is no such cancellations of electric fields. A clever calculation of the electrostatic potential of the infinity replicas should be performed in order to separate the potentials.
The difficulty now is that the replicated system
of just ions is no longer charge neutral, so that the electrostatic 
potential produced by the images of all the ions will diverge.  We will show,
however, that this divergence can be renormalized away, allowing us
to study a non-neutral periodic charged system. 

\begin{figure}[h]
\begin{center}
\includegraphics[scale=0.3]{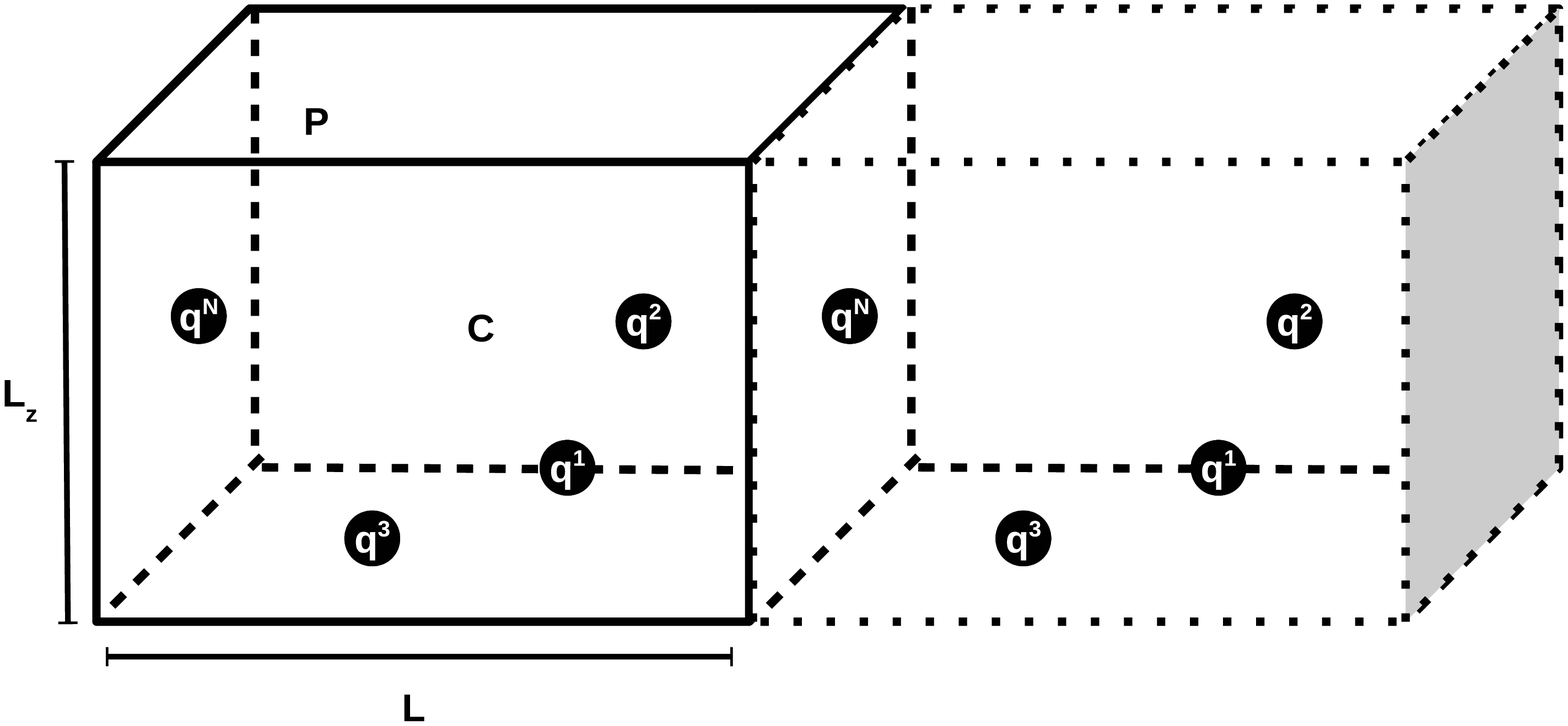}
\end{center}
\caption{The simulation box with randomly positioned charges and one of its replicas. The point C represents the center of the simulation box, where the origin is located, and P a random point.}
\label{fig2}
\end{figure}

Consider a system of particles of charges q$^j$ located at random positions ${\pmb r}^j$ inside a box of sides $L_x$, $L_y$ and $L_z$. The system in general is not charge neutral. Lets consider, without loss of generality, $L_x=L_y=L$. The system is now replicated infinitely in all directions. The replication vector is defined as ${\pmb r}_{ep}=(L n_1,L n_2,L_z n_3)$, where $n's$ are integers. In Fig.~\ref{fig2} we show the replicated system.
The electrostatic potential generated by the ions and all the images  at  a point P, located at some random position ${\pmb r}$ in the simulation box, can be written as
\begin{eqnarray}\label{elec_pot}
\phi({\pmb r})=\sum_{\pmb n}^{\infty}\sum_{j=1}^{N}\int\frac{\rho^j({\pmb s})}{\epsilon_w |{\pmb r}-{\pmb s}|}d^3{\pmb s} \ ,
\end{eqnarray}
where $\rho^j({\pmb s})=\text{q}^j \delta({\pmb s}-{\pmb r}^j-{\pmb r}_{ep})$ is the
charge density of q$^j$ and its replicas. The vector ${\pmb n}=(n_1,n_2,n_3)$ represent all the replicas, and the simulation box corresponds to $(0,0,0)$. The  3d Ewald summation~\cite{AlTi87} is a very efficient method for performing summation over all the replicas. The idea is to place a neutralizing Gaussianly distributed charge on top of each ion and then subtract the potential produced by the Gaussian charges  from the total potential. The fundamental observation is that if the charge of each ion is neutralized by the
Gaussian charge, the resulting potential will be short ranged and can be easily accounted for using simple periodic boundary conditions.  On the other hand, the potential of the Gaussian charges can be efficiently calculated using the Fourier 
representation of the charge density.  In fact the
distribution does not need to be Gaussian, but this is the most common choice~\cite{He81}.  

The electrostatic potential after adding and subtracting the Gaussian charges 
is
\begin{eqnarray}\label{elec_pot_E}
\phi({\pmb r})=\sum_{\pmb n}^{\infty}\sum_{j=1}^{N}\int\frac{\rho^j({\pmb s})-\rho^j_G({\pmb s})}{\epsilon_w |{\pmb r}-{\pmb s}|}d^3{\pmb s} + \nonumber \\
\sum_{\pmb n}^{\infty}\sum_{j=1}^{N}\int\frac{\rho^j_G({\pmb s})}{\epsilon_w |{\pmb r}-{\pmb s}|}d^3{\pmb s} \ ,
\end{eqnarray}
where $\rho^j_G({\pmb s})=\text{q}^j (\kappa_e^3/\sqrt{\pi^3})\exp{(-\kappa_e^2|{\pmb s}-{\pmb r}^j-{\pmb
r}_{ep}|^2)}$ and $\kappa_e$ is a dumping parameter. The potential can be written as
\begin{eqnarray}
\phi({\pmb r})=\sum_{\pmb n}^{\infty}\sum_{j=1}^{N}\text{q}^j\frac{\text{erf}(\kappa_e|{\pmb r}-{\pmb r}^j-{\pmb r}_{ep}|)}{\epsilon_w |{\pmb r}-{\pmb r}^j-{\pmb r}_{ep}|} + \nonumber \\
\sum_{\pmb n}^{\infty}\sum_{j=1}^{N}\text{q}^j\frac{\text{erfc}(\kappa_e|{\pmb r}-{\pmb r}^j-{\pmb r}_{ep}|)}{\epsilon_w |{\pmb r}-{\pmb r}^j-{\pmb r}_{ep}|} \ .
\end{eqnarray}
Using the Fourier transform the expression above can be written as
\begin{eqnarray}\label{elec_pot_E_f}
\phi({\pmb r})=\sum_{{\pmb k}={\pmb 0}}^{\infty}\sum_{j=1}^{N}\frac{4\pi \text{q}^j}{\epsilon_w V |{\pmb k}|^2}\exp{[-\frac{|{\pmb k}|^2}{4\kappa_e^2}+i{\pmb k}\cdot({\pmb r}-{\pmb r}^j)]} + \nonumber \\
\sum_{j=1}^{N}\text{q}^j\frac{\text{erfc}(\kappa_e|{\pmb r}-{\pmb r}^j|)}{\epsilon_w |{\pmb r}-{\pmb r}^j|} \ ,
\end{eqnarray}
where ${\pmb k}=(\frac{2\pi}{L}n_1,\frac{2\pi}{L}n_2,\frac{2\pi}{L_z}n_3)$.  In the second term of Eq.~\ref{elec_pot_E} we removed the summation, considering only the main box, ${\pmb n=}(0,0,0)$, with the usual periodic boundary condition. This is justified when $\kappa_e$ is sufficiently large, so that erfc$(\kappa_e |{\pmb r}|)$ decays rapidly, and the minimum image convention (periodic boundary condition) can be used. In practice we set $\kappa_e=5/L$, if $L<L_z$ or $\kappa_e=5/L_z$, if $L>L_z$. 

For ${\pmb k}=(0,0,0)$ the first term of Eq.~\ref{elec_pot_E_f} is singular. This singular term is discussed in a serie of publications~\cite{AlTi87,FrSm02,LaHu08,StTr11}. In order to treat it, some authors argue that we must consider the induced surface charge at the ``boundary of the infinity system" with the external medium. They argue that this term can be neglected if the exterior medium is a metal, a tinfoil boundary condition. At first, if we are dealing with an infinity system, there is not a ``boundary" defined. For a finite spherical system with an exterior medium of different dielectric constant, the boundary conditions can be satisfied if we consider, for example, image charges on there~\cite{DoBa11,DiDo12}. Lets consider it in more detail.  Neglecting the prefactors,
the ${\pmb k}=(0,0,0)$ term of the sum can be written as 
\begin{eqnarray}
\lim_{{\pmb k} \rightarrow 0}\sum_{j=1}^{N}\frac{q^j}{|{\pmb k}|^2}\exp{[-\frac{|{\pmb k}|^2}{4\kappa_e^2}]}\exp{[+i{\pmb k}\cdot({\pmb r}-{\pmb r}^j)]} \ .
\end{eqnarray}
Now, lets expand the exponentials and keep only the singular terms,
\begin{eqnarray}\label{pot}
\lim_{{\pmb k} \rightarrow 0}\sum_{j=1}^{N}\text{q}^j\frac{1}{|{\pmb k}|^2}-\sum_{j=1}^{N}\text{q}^j\frac{1}{4\kappa_e^2}+\nonumber \\
\lim_{{\pmb k} \rightarrow 0}\sum_{j=1}^{N}\text{q}^j\frac{i{\pmb k}\cdot({\pmb r}-{\pmb r}^j)}{|{\pmb k}|^2} -
\lim_{{\pmb k} \rightarrow 0}\sum_{j=1}^{N}\text{q}^j\dfrac{[{\pmb k}\cdot({\pmb r}-{\pmb r}^j)]^2}{2|{\pmb k}|^2} \ .
\end{eqnarray}

In a charge neutral system, $\sum_j \text{q}^j=0$, the first two terms are zero.  For a non-neutral system, however, they are infinite. On the other hand, they are 
independent of ${\pmb r}$, and can be renormalized away by simply redefining the zero of the potential. The third and fourth terms of expression (\ref{pot}) are position dependent and require greater care when calculating the limit ${\pmb k} \rightarrow 0$.
We first observe that the singular behavior of the ${\pmb k} \rightarrow 0$ is a
consequence of the large distance behavior of the lattice sum.  To properly account for this limit we rewrite the third and the fourth terms of expression (\ref{pot}) using the Dirac delta function.  
The third term can then be expressed as 
\begin{eqnarray}\label{s3}
S_3=\sum_{j=1}^{N}\text{q}^j\int_{-\infty}^{+\infty} \delta({\pmb k})\frac{i{\pmb k}\cdot({\pmb r}-{\pmb r}^j)}{|{\pmb k}|^2}d^3{\pmb k} \ ,
\end{eqnarray}
with the following representation of  $\delta({\pmb k})=\dfrac{1}{(2\pi)^3}\int_{-{\pmb H}}^{{\pmb H}} e^{i{\pmb k}\cdot{\pmb p}}d^3p$.

The limits of integration, $-{\pmb H}$ and ${\pmb H}$, where ${\pmb H}=(H_1,H_2,H_3)$ correspond to the way that the sums are performed in the real space. For example, if we replicate the simulation cell in a spherically symmetric fashion, then $H_1=lim_{m \rightarrow \infty} m L_x$, $H_2=lim_{m \rightarrow \infty} m L_y$, and $H_3=lim_{m \rightarrow \infty} m L_y$, that is
all sides diverge at the same rate.  On the other hand for a 
slab geometry $H_1$ and $H_2$ limits should go to infinity much faster than $H_3$.  In general it is
convenient to define $H_1=\alpha_1 L_c$, $H_2=\alpha_2 L_c$ and $H_3=\alpha_3 L_c$, where $L_c$ is some characteristic macroscopic length scale.   The ratio of  $\alpha's$ then corresponds to the aspect ratio of the macroscopic system, {\it i.e.} the simulation cell and all of its replicas. The integral over $p_1$, $p_2$, and $p_3$ 
can be performed explicitly yielding the following representation of the delta function,
\begin{eqnarray}
\delta({\pmb k})=\frac{1}{(2 \pi)^3}\prod_{i=1}^3\int_{-\alpha_i\frac{L_c}{2}}^{\alpha_i\frac{L_c}{2}} e^{i k_i p_i} dp_i=\frac{1}{ \pi^3}\prod_{i=1}^3\frac{\text{sin}(k_i\alpha_i L_c/2)}{k_i} \ .
\end{eqnarray}
This representation encodes the large distance behavior 
of the lattice sum and is at the heart of the singular behavior of ${\pmb k} \rightarrow 0$ limit. Eq. \ref{s3} can then be written as $S_3=\sum_{j=1}^{N} q_j {\pmb D} \cdot ({\pmb r}-{\pmb r}^j)$, where the components of the vector ${\pmb D}$ are,
\begin{eqnarray}
D_n= \frac{i}{\pi^3}\int_{-\infty}^{+\infty} \frac{ k_n}{|{\pmb k}|^2} \prod_{j=1}^3\frac{\text{sin}(k_j\alpha_j L_c/2)}{k_j}d^3{\pmb k}\ ,
\end{eqnarray}
which by symmetry integrate to  zero, $D_n=0$, so that $S_3=0$.

The fourth singular term of  expression (\ref{pot}) is
\begin{eqnarray}
S_4=-\sum_{j=1}^{N}\text{q}^j\int_{-\infty}^{+\infty}\delta({\pmb k})\dfrac{[{\pmb k}\cdot({\pmb r}-{\pmb r}^j)]^2}{2|{\pmb k}|^2}d^3{\pmb k} \ .
\end{eqnarray}
Again using the representation of the delta function it can be rewritten as 
\begin{eqnarray}
S_4=-\sum_{j=1}^{N}\frac{\text{q}^j}{2  \pi^3}\sum_{n=1}^3 B_n(r_n-r^j_n)^2 \ ,
\end{eqnarray}
where the index $n$ corresponds to the $x$, $y$, and $z$ components of the vector 
${\pmb r}$ and 
\begin{eqnarray}
B_n= \int_{-\infty}^{+\infty} d^3{\pmb k} \frac{ k_n^2}{|{\pmb k}|^2} \prod_{j=1}^3\frac{\text{sin}(k_j\alpha_j L_c/2)}{k_j}\ ,
\end{eqnarray}
Using the identity
\begin{eqnarray}
\frac{1}{{|\pmb k|}^2}=\int_0^\infty dt\ e^{-t {\pmb k}^2}\ ,
\end{eqnarray}
the coefficients $B_n$ can be simplified to~\cite{Sm81}
\begin{eqnarray}
B_1=\dfrac{\pi^{\frac{5}{2}}}{2}\int_{0}^{+\infty} \dfrac{\alpha_{13} e^{-\frac{\alpha_{13}^2}{4 t}}\text{erf}(\frac{\alpha_{23}}{2\sqrt{t}})\text{erf}(\frac{1}{2\sqrt{t}})}{t^{\frac{3}{2}}} dt \ ,
\end{eqnarray}
\begin{eqnarray}
B_2=\dfrac{\pi^{\frac{5}{2}}}{2}\int_{0}^{+\infty} \dfrac{\alpha_{23} e^{-\frac{\alpha_{23}^2}{4 t}}\text{erf}(\frac{\alpha_{13}}{2\sqrt{t}})\text{erf}(\frac{1}{2\sqrt{t}})}{t^{\frac{3}{2}}} dt \ ,
\end{eqnarray}
\begin{eqnarray}
B_3=\dfrac{\pi^{\frac{5}{2}}}{2}\int_{0}^{+\infty} \dfrac{e^{-\frac{1}{4 t}}\text{erf}(\frac{\alpha_{13}}{2\sqrt{t}})\text{erf}(\frac{\alpha_{23}}{2\sqrt{t}})}{t^{\frac{3}{2}}} dt \ .
\end{eqnarray}
where $\alpha_{ij}=\alpha_i/\alpha_j$ are the aspect ratios of the macroscopic
system.  The coefficients $B_n$ can now be easily calculated using numerical integration.  For a spherically symmetric summation of replicas the 
aspect ratios are $\alpha_{13}=L_x/L_z$ and $\alpha_{23}=L_y/L_z$. On the other hand,
for a planewise summation of a slab geometry,  $\alpha_{13}\rightarrow\infty$ and $\alpha_{23}\rightarrow\infty$. In this case
the integrals can be performed explicitly~\cite{Sm81} yielding $B_1=B_2=0$, and $B_3=\pi^3$.

Separating the ${\pmb k}={\pmb 0}$ term from the ${\pmb k}$-vector summation, Eq.~(\ref{elec_pot_E_f}) can now be rewritten as
\begin{eqnarray}\label{elec_pot_E_new}
\Delta \phi({\pmb r})=\sum_{{\pmb k}\neq{\pmb 0}}^{\infty}\sum_{j=1}^{N}\frac{4\pi \text{q}^j}{\epsilon_w V |{\pmb k}|^2}\exp{[-\frac{|{\pmb k}|^2}{4\kappa_e^2}+i{\pmb k}\cdot({\pmb r}-{\pmb r}^j)]} + \nonumber \\
-\sum_{j=1}^{N}\sum_{n=1}^3 \dfrac{2\text{q}^j}{\epsilon_w V \pi^2}B_n(r_n-r^j_n)^2 + \hspace{1cm} \nonumber \\
\sum_{j=1}^{N}\text{q}^j\frac{\text{erfc}(\kappa_e|{\pmb r}-{\pmb r}^j|)}{\epsilon_w |{\pmb r}-{\pmb r}^j|} \ , \hspace{1cm}
\end{eqnarray}
where $\Delta$ corresponds to the renormalization of the potential in Eq.~(\ref{elec_pot_E_f}). As a test of the modified Ewald summation formula for a non-neutral system, Eq. (\ref{elec_pot_E_new}), we
calculate the electrostatic potential difference between a random position ${\pmb r}$ and the center of the simulation box, ${\pmb 0}$.  Note that although the 
electrostatic potential is divergent for a non-neutral periodic system, the potential difference is well defined.  We calculate the renormalized electrostatic potential produced by the two charges  $q_1=q_2=|e|$, where $e$ is the electron charge, located at random positions. We set $L_x=L_y=L=1~$\AA\ and $L_z=2~$\AA. The spherical replication of the rectangular box will result in an infinite system with an aspect ratio of $\alpha_{13}=1/2$ and $\alpha_{23}=1/2$, leading to  parameters $B_1=B_2=13.5158$ and $B_3=3.9746$. Using Eq.~\ref{elec_pot_E_new}, we find the converged value  $\Delta\phi=\phi({\pmb r})-\phi({\pmb 0})$, to $2-$decimal place accuracy, using $\approx~250$ k-vectors spherically summed. In real space using the explicit summation, Eq.~\ref{elec_pot}, we find exactly the same converged value for $\Delta\phi({\pmb r})$, the convergence, however, is much slower, so that to get a $2-$decimal place accuracy requires summation of over $\approx~19700$ n-vectors. Our findings agree with the results of Nymand and Linse~\cite{NyLi00}, which compared the potentials for an anisotropic neutral system, see Table 1 of Ref.~\cite{NyLi00}. For isotropic simulations, the energy related with the singular term can be very small on average. This can explain some results found in literature using ``tinfoil" boundary conditions. However, in order to corretly describe an infinite electrostatic system, also anisotropic one, the singular term is important and it is not related with boundary conditions, only with the singular term.

In the slab geometry we want to calculate the potential difference when the simulation box is replicated in the $x$ and $y$ directions only. Again we 
will use the modified 3d Ewald summation given by Eq. (\ref{elec_pot_E_new}).  This means that the box will be
replicated in all 3 dimensions. However, the replication in the $x$ and $y$ directions 
should be performed at a rate much faster than in the $z$ direction.  This leads to 
$B_1=B_2=0$, and $B_3=\pi^3$ and  Eq. (\ref{elec_pot_E_new}) becomes 
\begin{eqnarray}\label{elec_pot_E_new2}
\Delta \phi({\pmb r})=\sum_{{\pmb k}\neq{\pmb 0}}^{\infty}\sum_{j=1}^{N}\frac{4\pi \text{q}^j}{\epsilon_w V |{\pmb k}|^2}\exp{[-\frac{|{\pmb k}|^2}{4\kappa_e^2}+i{\pmb k}\cdot({\pmb r}-{\pmb r}^j)]} + \nonumber \\
-\sum_{j=1}^{N}\frac{2\pi \text{q}^j}{\epsilon_w V}(r_3-r_3^j)^2 + \hspace{1cm} \nonumber \\
\sum_{j=1}^{N}\text{q}^j\frac{\text{erfc}(\kappa_e|{\pmb r}-{\pmb r}^j|)}{\epsilon_w |{\pmb r}-{\pmb r}^j|} \ . \hspace{1cm}
\end{eqnarray}

Even though the contribution from the $z$-directional replicas is much smaller than
from the $x$ and $y$ directional replicas, it is not negligible.  
In order to diminish the impact of $z$-replicas on the electrostatic potential, we
must leave a sufficiently large vacuum region in the $z$-direction.  
To test Eq. (\ref {elec_pot_E_new2}) for a slab geometry, 
we study the same 2 particle  system discussed earlier.
Using Eq.~\ref{elec_pot} we can explicitly calculate the potential difference $\Delta\phi=\phi({\pmb r})-\phi({\pmb 0})$, when the simulation cell is replicated
{\bf only} in the $x$ and $y$ directions, ${\pmb n}=(n_x,n_y,0)$.  
The convergence is very slow
requiring values of $2.5\times10^6$ replicas to get an accuracy of $2-$decimal places.

To diminish the interaction with $z$-directional replicas, in order to use 
Eq. (\ref {elec_pot_E_new2}) for a slab geometry, we restrict 
positions of the charges and the vector ${\pmb r}$ to the region $-\frac{L_z}{4}<z<\frac{L_z}{4}$ in the simulation cell, leaving the regions $-\frac{L_z}{2}<z<-\frac{L_z}{4}$ and $\frac{L_z}{4}<z<\frac{L_z}{2}$ empty. The calculated electrostatic potential 
difference is exactly the same as found using the real-space lattice summation.  
The same  $2-$decimal point accuracy, however, is achieved with only 
$\approx~630$ k-vectors. 

The renormalized electrostatic energy for a non-neutral slab  system can now be calculated as $E=\dfrac{1}{2}\sum_{i=1}^N\text{q}^i\Delta \phi({\pmb r}^ i)$,
\begin{eqnarray}\label{ener}
E=\sum_{{\pmb k}\neq{\pmb 0}}^{\infty}\frac{2\pi}{\epsilon_w V |{\pmb k}|^2}\exp{[-\frac{|{\pmb k}|^2}{4\kappa_e^2}]}[A({\pmb k})^2+B({\pmb k})^2] + \nonumber \\
\frac{2\pi}{\epsilon_w V}[M_z^2-Q_tG_z]
+ \hspace{1cm} \nonumber \\
\dfrac{1}{2}\sum_{i \ne j}^N\text{q}^i\text{q}^j\frac{\text{erfc}(\kappa_e|{\pmb r}^i-{\pmb r}^j|)}{\epsilon_w |{\pmb r}^i-{\pmb r}^j|} \ , \hspace{1cm}
\end{eqnarray}
where
\begin{eqnarray}
A({\pmb k})=\sum_{i=1}^N \text{q}^i\text{cos}({\pmb k}\cdot{\pmb r}^i) \ , \nonumber \\
B({\pmb k})=-\sum_{i=1}^N \text{q}^i\text{sin}({\pmb k}\cdot{\pmb r}^i) \ , \nonumber \\
M_z=\sum_{i=1}^N \text{q}^i r_3^i \ , \nonumber \\
Q_t=\sum_{i=1}^N \text{q}^i \ , \nonumber \\
G_z=\sum_{i=1}^N \text{q}^i (r_3^i)^2 \ .
\end{eqnarray}
For a neutral system, $Q_t=0$, and we recover the earlier expression for the electrostatic energy~\cite{YeBe99}.

\begin{figure}[t]
\begin{center}
\includegraphics[scale=0.6]{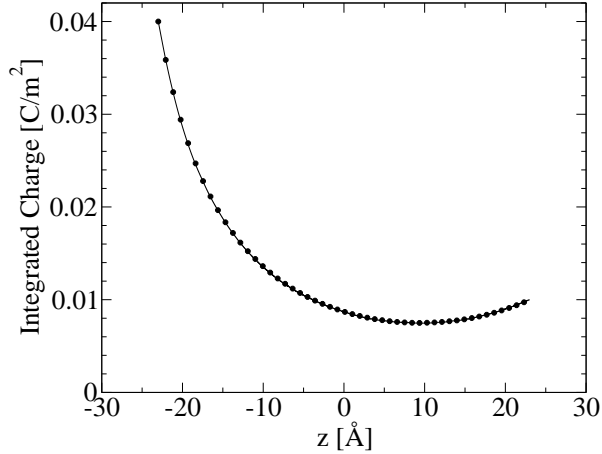}
\end{center}
\caption{Integrated charge between the plates. Symbols represent the calculation using the modified (non-neutral) 3d Ewald approach, 
while line, the traditional method~\cite{YeBe99}. The difference is imperceptible.}
\label{fig3}
\end{figure}

We now apply the method developed above to a system of electrolyte confined between two charged plates. We set $L=179~$\AA\ and $L_z=400~$\AA. The ionic radius is $2~$\AA, while the separation between plates is $50~$\AA. The number of k-vectors is around $300$. The equilibration is achieved with $1\times10^6$ MC steps, while the density profiles are obtained with $20000$ samples, each saved after $100$  particle trial moves. As a first example, we set $\sigma_1=0.04~$C$/$m$^2$ and $\sigma_2=-0.01~$C$/$m$^2$. For plate 1 we have $80$ counterions of charge $-|e|$, while for plate 2, we have $20$ counterions of charge $|e|$. In the MC Metropolis algorithm we use the energy expression Eq.~\ref{ener}, for the $N_c=100$ ions, and the electrostatic energy of interaction between ions and the charged plates,
\begin{eqnarray}\label{ep}
E_p=\dfrac{2\pi}{\epsilon_w}\sum_{i=1}^{N_c}(\sigma_2-\sigma_1)r_3^i\text{q}^i \ .
\end{eqnarray}

\begin{figure}[h]
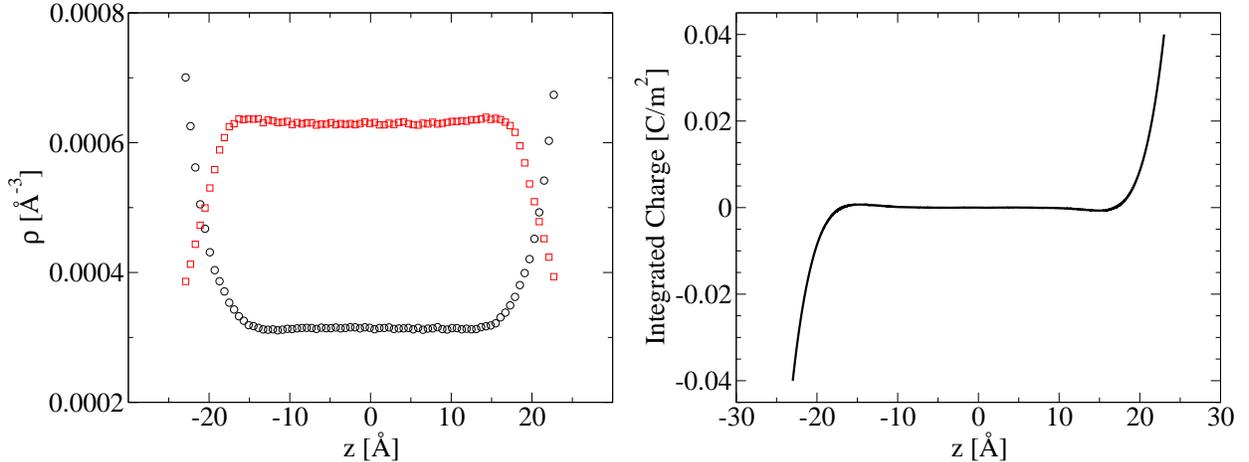

\begin{center}
\includegraphics[scale=0.6]{fig4a.eps}\vspace{0.3cm}\hspace{0.1cm}
\includegraphics[scale=0.6]{fig4b.eps}\vspace{0.3cm}\hspace{0.1cm}
\end{center}
\caption{(a) Density profiles of ions between equally charged plates for 2:1 salt - circles represent positive ions, while squares negative ones. (b) The integrated charge.}
\label{fig4}
\end{figure}
\begin{figure}[h]
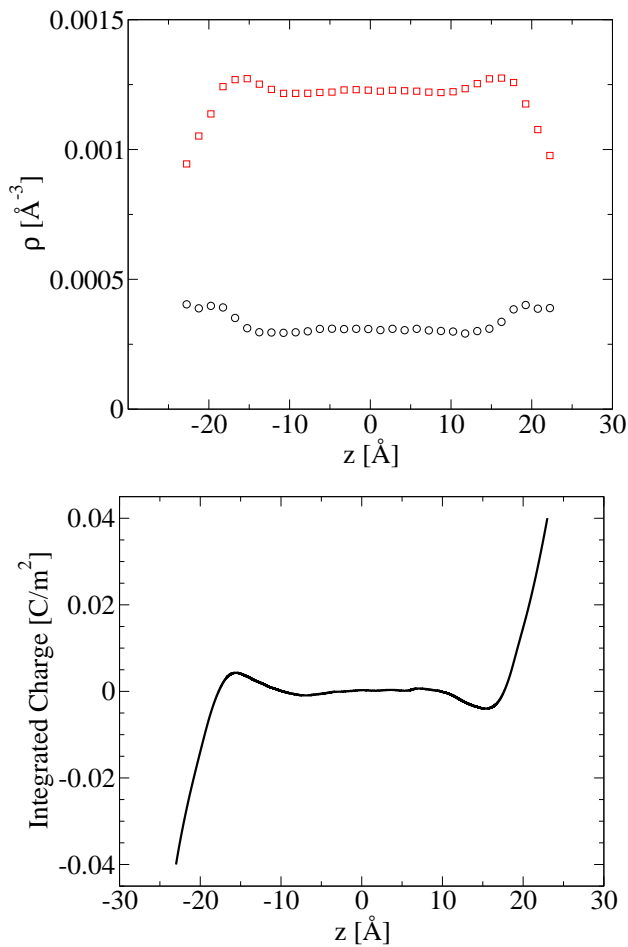

\begin{center}
\includegraphics[scale=0.6]{fig5a.eps}\vspace{0.3cm}\hspace{0.1cm}
\includegraphics[scale=0.6]{fig5b.eps}\vspace{0.3cm}\hspace{0.1cm}
\end{center}
\caption{(a) Density profiles of ions between equally charged plates for 4:1 salt - circles represent positive ions, while squares negative ones. (b) The integrated charge.}
\label{fig5}
\end{figure}

To appreciate the power of the present method we compare it with the usual algorithm in which the surface charge is represented by 256 uniformly distributed point particles~\cite{YeBe99}. In this case we 
use Eq.~\ref{ener} for a neutral system, considering all charged particles, including the ones on the plate surface, $Q_t=0$. The result is shown in Fig.~\ref{fig3} and is indistinguishable from the non-neutral simulation method developed in the present paper.  The gain in the simulation time is very substantial --- a traditional simulation method took $20$ times more CPU time than the algorithm developed in the present paper. Next we apply the new simulation method to the case of  $\sigma_1=\sigma_2$. This situation is particularly relevant for
studying colloidal stability with the help of Derjaguin approximation~\cite{Ru89}. We consider $500~$mM of 2:1 and 4:1 dissociated salts between  charged walls. The electric fields produced by the plates cancel out. Therefore, the simulation is performed only with the Eq.~\ref{ener} -- we do not need to take into account the plates in the calculations, except in order to obtain the number of plate counterions. For $\sigma_1=\sigma_2=-0.04~$C$/$m$^2$, we have $80$ counterions of charge $2|e|$, for 2:1 case and \bf $40$ counterions of charge $4|e|$, for 4:1 case. Using the same $L=179~$\AA$\,$ and $L_z=400~$\AA, the ionic profiles and integrated charges are shown in Fig.~\ref{fig4} and Fig.~\ref{fig5}. Observe that for 4:1 salt the inversion of charge is much more important than 2:1 case. A study of such strongly concentrated systems are not very practical with other simulation methods.

\section{Conclusions}
We have developed a new approach for simulating electrolytes in a confined slab
geometry. Our algorithm relies on 3d Ewald summation to properly account for the long range Coulomb interaction between the ions and the charged surfaces.   
The optimal performance of the method is 
achieved by separating the electrostatic potential produced by the  charged walls from the potential produced by the electrolyte. 
The fundamental observation that we make is that the electrostatic potential produced by the 3d periodic images of the plates has a simple linear form, with the electric field produced by the transverse images of the charged plates canceling out. This observation suggests that the ions and the charged surfaces can be treated separately. The difficulty, however, is that the system of only ions no longer respects the charge neutrality,
with its electrostatic energy diverging.  Nevertheless, we show that a simple renormalization of the electrostatic potential cures the divergence, allowing us to consider a non-neutral system of ions moving in the field produced by the charged plates.  This approach leads to a dramatic speed up of simulations of Coulomb systems confined between charged walls. The simulations can be made  to run even faster by adopting a Particle-Particle Particle-Mesh ($P^3M$) approach. Such improvement would allow to use our algorithm for studying  all atom large scale simulations of liquid-liquid/vapour interfaces~\cite{KuMu06,EgSi08}.
Finally, inclusion of  dielectric discontinuities can be easily implemented in the present method using image charges~\cite{DoLe15}. In the original method for dielectric walls~\cite{DoLe15}, only the correction part of energy, similar to the central term of Eq.~\ref{ener} of this manuscript, should be rederived because a charge neutrality condition is considered.

\section{Acknowledgments}
This work was partially supported by the CNPq, INCT-FCx, and by the US-AFOSR under the grant FA9550-12-1-0438.

\bibliography{ref.bib}

\end{document}